\documentclass[sigconf, nonacm]{acmart}

\renewcommand\footnotetextcopyrightpermission[1]{} 
\AtBeginDocument{%
  }

\usepackage{soul}
\usepackage{booktabs}

\copyrightyear{2026}
\acmYear{2026}
\setcopyright{cc}
\setcctype{by}

\newif\ifsubmit
\submitfalse
\ifsubmit
    \newcommand{\changes}[1]{}

\else
    \newcommand{\changes}[1]{{\leavevmode\color[rgb]{0.0, 0.0, 0.0}{#1}}}
\fi

\begin{document}

\title{Compos3D: Interactive Part-Based Composition for Creative Control in Generative 3D Models}

\author{Faraz Faruqi}
\email{faraz.faruqi@autodesk.com}
\affiliation{%
  \institution{Autodesk Research}
  \city{Boston}
  \state{MA}
  \country{USA}
}

\author{Sean J. Liu}
\email{sean.liu@autodesk.com}
\affiliation{%
  \institution{Autodesk Research}
  \city{San Francisco}
  \state{California}
  \country{USA}
}

\author{George Fitzmaurice}
\email{George.Fitzmaurice@autodesk.com}
\affiliation{%
  \institution{Autodesk Research}
  \city{Toronto}
  \state{Ontario}
  \country{Canada}
}

\author{Justin Matejka}
\email{Justin.Matejka@autodesk.com}
\affiliation{%
  \institution{Autodesk Research}
  \city{Toronto}
  \state{Ontario}
  \country{Canada}
}

\renewcommand{\shortauthors}{Faruqi et al.}



\begin{abstract}

While generative AI has unlocked new opportunities for 3D content creation, current workflows often rely on multiple regenerations, which provides limited control and unpredictable outcomes. We present Compos3D, a system that introduces a compositional workflow for generative 3D modeling through remixing. Instead of repeatedly regenerating models, users generate multiple candidates from text or image prompts, select parts of interest via 2D image regions or 3D mesh segments, and assemble them into a coherent design. The system synthesizes these compositions into a refined 3D model, preserving high-level intent while resolving low-level geometry. 
By decoupling the 3D generation workflow into two stages--generating intermediate candidates and subsequently remixing them into the final design--our approach gives users finer-grained control over the final output.
To evaluate this approach, we conducted a controlled user study comparing remixing and regeneration workflows across both 2D and 3D modalities. Results show that the remixing workflow provides participants with greater creative control, stronger alignment with their intent, and higher satisfaction. We conclude with design recommendations for future AI-assisted 3D modeling workflows.

\end{abstract}

\begin{teaserfigure}
  \includegraphics[width=\textwidth]{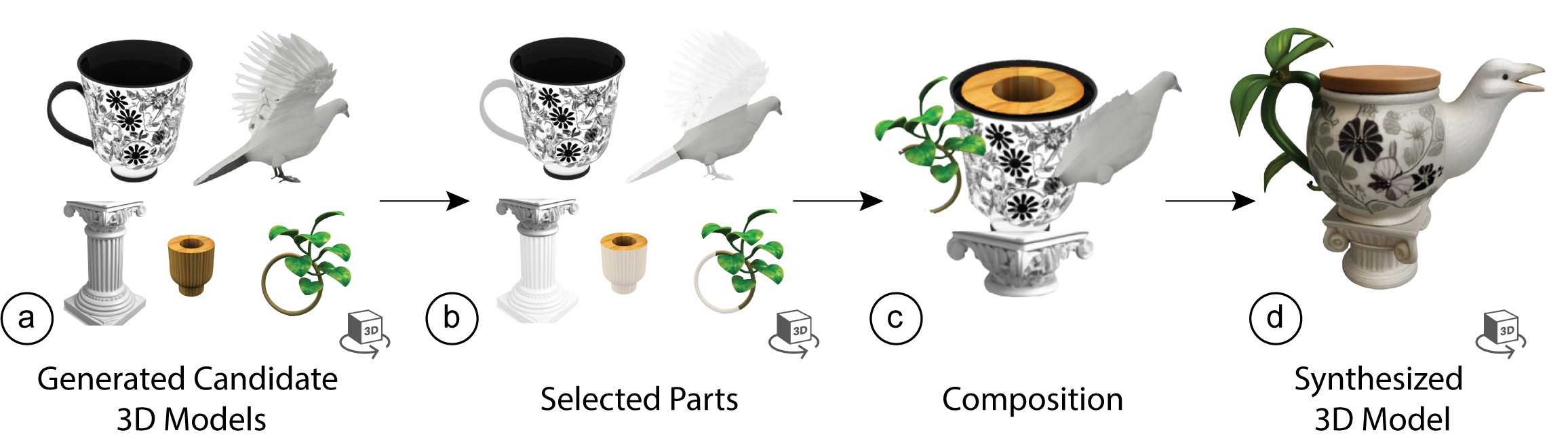}
  \caption{Compos3D introduces a compositional remixing workflow for generative 3D modeling. Users (a)~generate multiple candidate 3D models from text or image prompts, (b)~select desired parts from different models, and (c)~rearrange the parts into their desired (rough) composition. (d)~The system then synthesizes a coherent 3D model that integrates the chosen parts into a unified design. 
  }
  \label{fig:teaser}
\end{teaserfigure}


\maketitle

\section{Introduction}

The rapid growth in 3D Generative AI tools has enabled users to create increasingly complex 3D models from simple text or image prompts, opening new opportunities for creative design where novice creators can rapidly realize their ideas without learning complex 3D modeling techniques. However, most 3D generation systems currently operate as one-shot black boxes: users describe an object, receive one or several 3D models, but have little ability to manipulate the results beyond the prompt. When users want to incorporate specific changes, such as 
preserving a particular element while modifying the rest, they face a frustrating ``\textit{dice roll}'' scenario, repeatedly regenerating with modified prompts and hoping the system preserves what they liked while producing something new. Each regeneration risks losing desired features, and the process provides no mechanism for users to directly express their design intent beyond words.

In contrast, creative practice across domains has long relied on a different approach: composing new works by selecting and recombining parts from multiple sources. Designers build moodboards to externalize ideas that are hard to express verbally~\cite{cassidy2008moodboard_ref}. In generative art and creative coding, practitioners fork, remix, and recombine outputs to explore new directions~\cite{angert2023spellburst, subbaraman2023forking}. 
Research across various domains consistently shows that such compositional workflows where users demonstrate intent by assembling parts rather than describing a final result, foster deeper creative engagement, stronger alignment with design goals, and greater feelings of ownership~\cite{brade2023promptify, masson2024directgpt, peng2025fuseain}. 

\changes{
While recent work has enabled localized refinement of individual 3D~models through text-guided editing and style transfer~\cite{lu2024advances, dreameditor2023, tipeditor2024}, 
the regeneration workflow largely remains the same:
users must accept or reject whole models rather than treat them as raw material for subsequent design. This also limits how users can leverage the rich diversity of the model’s generative space.

}

We introduce \textit{Compos3D}, a system that brings compositional remixing to generative 3D modeling. Rather than treating generated models as indivisible outputs to accept or discard, Compos3D treats them as intermediate source material from which users can extract, arrange, and recombine parts. We define remixing as the practice of taking elements from disparate designs and combining them into a refined model that preserves and integrates distinct features from multiple sources. In our workflow, users first use an underlying AI model to generate multiple 3D candidates from text or image prompts, then select parts of interest, such as the wings from one model and the body of another, as 2D image regions or 3D mesh segments, and arrange them into a rough composition. The system synthesizes these compositions into a coherent 3D model, preserving the user's high-level design intent while resolving low-level geometry and topology. 

By decoupling 3D generation into a two-stage process: first producing intermediate candidate outputs, then remixing them into a final design, our approach gives users more precise control while leveraging the rich diversity of the model's generative space. 
For part selection and composition, Compos3D supports both 2D (image-based) and 3D (mesh-based) controls, as each offers complementary affordances. 2D interactions provide speed and flexibility for rapid collage-making, drawing motivation from familiar image editing processes~\cite{liu2025generative}, while 3D interactions afford precise spatial control over positioning and proportion. 
 
To evaluate our approach, we conducted a controlled user study with eight novice 3D modelers, comparing Compos3D's remixing workflow against a regeneration workflow, in which users iteratively refine a single model through local region selection and text-based instructions. We chose this baseline as it is the predominant interaction paradigm in current 3D generative tools. Our goal was to understand whether compositional remixing offers distinct benefits in creative control, alignment with design intent, and user satisfaction with the outcome, over the regeneration baseline. 
We also identify any trade-offs between 2D and 3D interaction modalities for such a workflow. 
Results show that remixing gave participants greater creative control, stronger alignment with their intent, and higher satisfaction, while requiring modestly more effort. We conclude with design recommendations for integrating compositional workflows into future AI-assisted 3D modeling tools.
 
In summary, our work contributes:
 
\begin{enumerate}
    \item A system for multi-modal remixing of generative 3D models through image-based and mesh-based part selection, composition, and AI-assisted synthesis;
    \item A controlled user study comparing remixing and regeneration workflows for creative 3D design;
    \item A comparison of 2D and 3D interaction modalities for compositional remixing;
    \item Design recommendations for integrating compositional workflows into future 3D generative tools.
\end{enumerate}

\section{Related Work}


We situate our work at the intersection of human-computer interaction, generative AI, and computational design. Specifically, we draw on prior research in three key areas: (1) advances in 3D generative AI, (2) current support tools for 3D modeling, and (3) remixing as a creativity-support paradigm.

\subsection{3D Generative AI Workflows}

Reconstructing 3D models from a single monocular image or a text description is a longstanding challenge in computer vision. Recent machine learning methods address this problem by learning priors over large collections of 3D shapes to infer the full geometry of an object from a single image. Several approaches have been proposed for this problem, including but not limited to GANs~\cite{cai2020learning, gao2022get3d}, flow networks~\cite{klokov2020discrete, yang2019pointflow}, VAEs~\cite{mittal2022autosdf, wu2019sagnet}, and diffusion models~\cite{chou2023diffusion, jun2023shap_e}. The recent Large-Reconstruction Model (LRM) architecture~\cite{hong2023lrm}-based methods~\cite{xu2024instantmesh, tang2024lgm, boss2024sf3d} have demonstrated generalizable and high-quality 3D reconstruction.

These generative models have enabled the development of support tools for creative exploration and rapid prototyping~\cite{shen2024neural}. 3DALL-E~\cite{liu20233dall} integrates text-to-image AI into CAD software to provide designers with image-based inspiration. 
VRCopilot~\cite{zhang2024vrcopilot} proposed a mixed-initiative VR authoring system that integrates generative AI with multimodal interactions and intermediate wireframe representations. Style2Fab~\cite{faruqi2023style2fab} introduces a method to selectively personalize 3D models while preserving their affordances for physical use, while TactStyle~\cite{faruqi2025tactstyle} enabled the stylization of 3D models using image prompts while controlling their visual and tactile properties. 

While these advances have lowered the barrier for users to create unique and high-quality 3D models, 
the underlying workflow largely remains single-shot, where users generate and evaluate complete outputs in a single step.
To incorporate changes, users typically regenerate entire models, with limited ability to preserve desired features or combine elements across outputs. In contrast, image-generation research has shown that compositional interfaces such as Generative Photomontage~\cite{liu2025generative} give users more concrete control for creative exploration. Inspired by these approaches, we explore how similar compositional workflows can be brought into 3D generative modeling, enabling novices to remix parts across multiple outputs rather than rely solely on trial-and-error prompting.

\subsection{3D Modeling Support Tools for Novices}

3D modeling requires both design expertise and proficiency with complex CAD software, making it difficult for novices to translate intent into design. 

HCI researchers have addressed these barriers with tools that scaffold the modeling process and lower the threshold for participation.
For instance, Meshmixer~\cite{schmidt2010meshmixer} lets users combine parts extracted from a library of meshes, while PARTS~\cite{hofmann2018greater} provides a method for designers to label their geometry with custom programs, making the models more reusable by novice makers. Alternatively, AutoConnect~\cite{koyama2015autoconnect} allows users to connect two meshes without modifying them by creating a connector as a new mesh. Grafter~\cite{roumen2018grafter} creates new mechanisms from existing parts by scaling the meshes to fit the mechanisms without affecting their functionality. Attribit~\cite{chaudhuri2013attribit} enables replacement using semantic attributes, and parametric design systems~\cite{shugrina2015fab, schulz2014design, veuskens2020coda} provide interactive customization while preserving validity and manufacturability. 

While these 3D modeling support tools enable novice users to personalize existing meshes or models, they are either limited to a static library of models, or only allow limited customization. Generative AI methods offer a way beyond static libraries and limited parametric customization; however, current 3D tools still rely on one-shot generation or regeneration, limiting their usability. Our work addresses this gap by integrating remixing into 3D generative workflows.


\subsection{Human-in-the-Loop Customization with Generative Models}  

Across domains, HCI researchers have designed systems that situate generative models in human-in-the-loop workflows, giving users ways to steer, remix, and reuse outputs rather than rely on one-shot prompts. Previous studies on non-experts with AI-based models~\cite{zamfirescu2023johnny, subramonyam2023bridging} highlight why such interfaces are needed: novices struggle with underspecified prompts, opaque model reasoning, and lack of feedback. Interactive systems that afford multi-modal interaction and provide intuitive and controllable feedback mechanisms mitigate these challenges by scaffolding exploration, making intermediate steps visible, and enabling correction or recombination.

In creative coding, tools such as Spellburst~\cite{angert2023spellburst}, and XCreation~\cite{yan2023xcreation} provide node or graph-based interfaces to compose multimodal generative pipelines. WorldSmith~\cite{dang2023worldsmith} and AngleKindling~\cite{petridis2023anglekindling} let writers fork, merge, and refine prompts for narrative and journalistic content. In visual art and design, systems such as PromptPaint~\cite{chung2023promptpaint}, Promptify~\cite{brade2023promptify}, and PromptCharm~\cite{wang2024promptcharm} allow localized steering of image generation through brush strokes, prompt suggestions, attention-map reweighting, or direct manipulation. DirectGPT~\cite{masson2024directgpt} generalizes this idea to text, code, and graphics, translating GUI actions into prompt edits that embody direct manipulation principles. 

Other work has emphasized organizing and reusing outputs. HistoryPalette~\cite{benharrak2025historypalette} and DreamSheets~\cite{almeda2024prompting} capture and cluster generations, supporting revision and exploration of alternatives. Studies of example galleries~\cite{yang2024considering, lee2010designing} highlight how curating and reusing design examples is integral to creative practice. Research on remixing communities~\cite{subbaraman2023forking} shows that branching, forking, and reuse are central to creative culture, aligning with systems that scaffold iterative exploration.  
In the LLM domain, systems like EvalLLM~\cite{kim2024evallm}, AIChains~\cite{wu2022ai}, and Sensecape~\cite{suh2023sensecape} allow breaking complex tasks into smaller steps, generate multiple results for different prompts, and evaluate them on user-defined criteria.
 
Together, this body of work shows that compositional interactive workflows are useful across nearly all domains of generative AI: from writing to art, to programming. They suggest that giving users explicit ways to preserve, recombine, and organize generative outputs fosters creativity, control, and ownership. We extend this line of work into 3D generative modeling by allowing users to combine parts across multiple outputs while retaining precise control over how they are integrated.

\section{System Design and User Interface} 

Existing generative 3D tools often treat creation as a one-shot process: a prompt yields a complete model, which users can only adjust through global style edits or repeated regeneration. This limits fine-grained control and prevents designers from reusing desirable elements across different outputs. Compos3D addresses this gap by enabling part-based remixing: users can extract meaningful components from multiple generated or uploaded models and assemble them into new hybrid designs. Inspired by creative practices like collaging and kitbashing, our system treats intermediate generations as raw material for composition rather than discarded attempts.

To support this workflow, Compos3D provides a unified pipeline with three phases: Part Selection, Composition, and Synthesis, available in both 2D (image-based) and 3D (mesh-based) modalities. In Part Selection, users segment and select desired parts from candidate models; in Composition, they arrange these parts into rough collages on a 2D canvas or within a 3D scene; and in Synthesis, the system refines these collages into coherent, high-quality 3D models that preserve the user’s design intent. In the next sections, we describe each step in more detail.

\begin{figure}
    \centering
    \includegraphics[width=\linewidth]{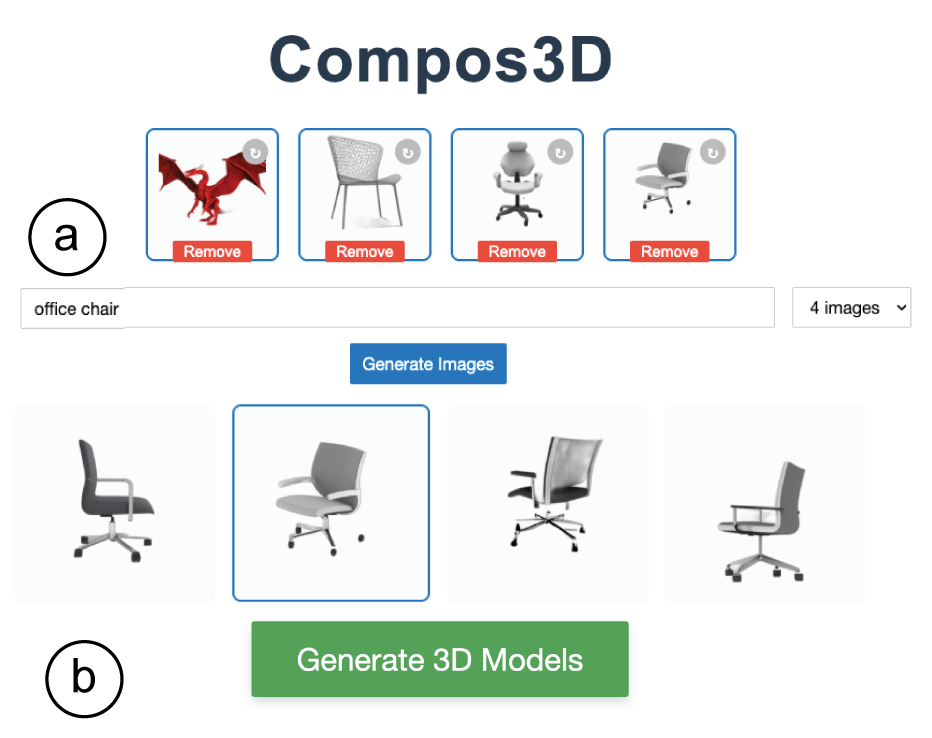}
    \caption{Compos3D 3D Model Generation Interface. (a)~Images selected by the user for 3D generation, displayed in a carousel. (b)~Candidate images generated from a text prompt (e.g., ``office chair''), from which users select images of interest. Once satisfied, the user clicks `Generate 3D Models' to convert the selected images into 3D models using an image-to-3D generative AI model.}
 \label{fig:user_interface-generation}

\end{figure}

\begin{figure}
    \centering
    \includegraphics[width=\linewidth]{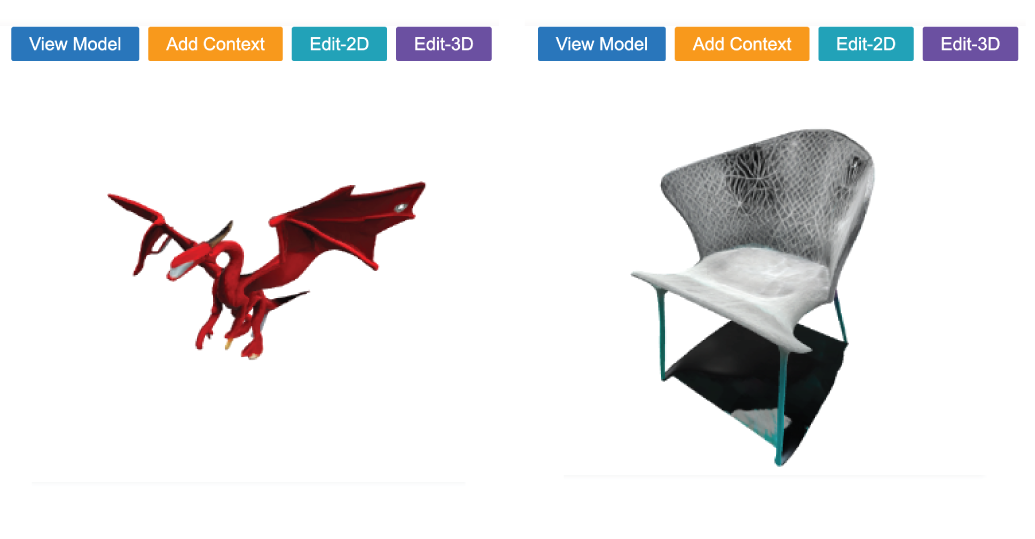}

    \caption{Generated 3D model viewer. Each generated model is displayed with options to inspect it in a larger 3D viewport (\textit{View Model}) and add it to the remix workspace (\textit{Add Context}). For comparison with the regeneration workflow (Section \ref{sec:user_study}), we allow users to regenerate the models (\textit{Edit 2D}, \textit{Edit 3D}).
    }

 \label{fig:user_interface-generation-viewer}

\end{figure}

\subsection{Initial Model Generation and Remix Context}
\label{sec:initial_model_gen}
Users begin with generating initial candidate 3D models using either text or image prompts (Fig.~\ref{fig:user_interface-generation}a). If given a text prompt, such as “office chair”, Compos3D first generates a diverse set of images with a text-to-image model, from which users select images of interest into a carousel. Our system then converts the selected images into candidate 3D models with an image-to-3D model~\cite{trellis} (Fig.~\ref{fig:user_interface-generation-viewer}). If given an image prompt, Compos3D directly converts it to candidate 3D models with the image-to-3D model. The resulting 3D models are rendered in the interface for the user's inspection. The user can open them in a larger viewer and view the geometry from multiple angles. Then, users can select models to remix by adding promising designs into a Remix Context (Fig.~\ref{fig:user_interface-generation-viewer}). This curated pool of 3D models serves as the working set from which parts can later be extracted and recombined.

In addition to generated outputs, users may also import their own 2D or 3D assets, allowing remixing with elements drawn from existing libraries or real-world artifacts.

\begin{figure}
    \centering
    \includegraphics[width=\linewidth]{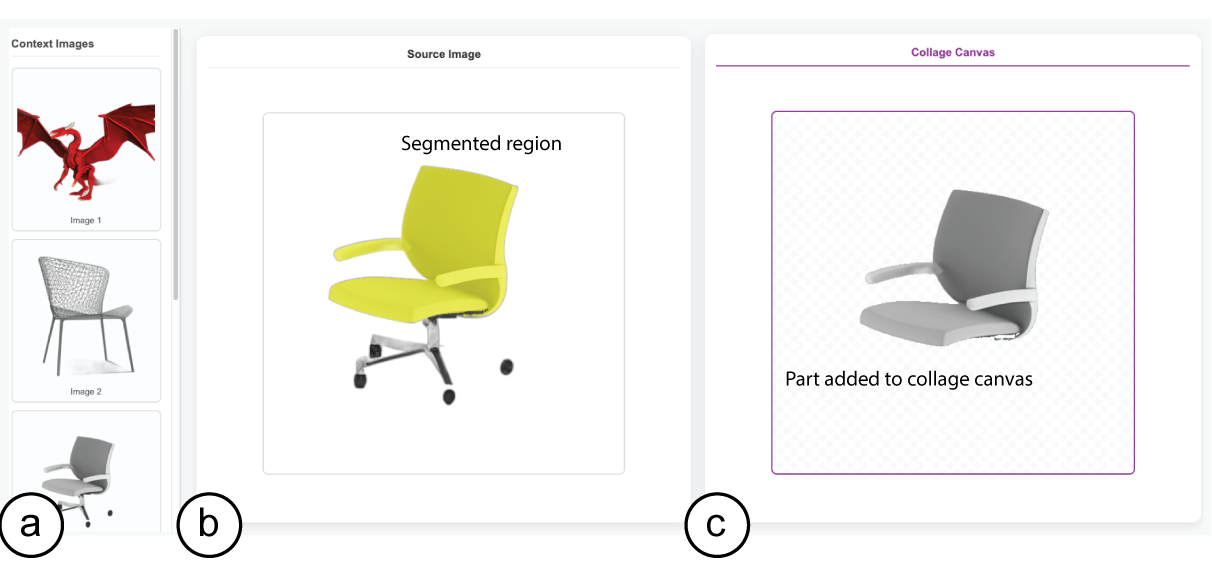}
\caption{2D Remixing Workspace. (a)~Source images from the remix context. (b)~The user selects a region of interest from a source image using click-based segmentation. (c)~The extracted segment is placed on the collage canvas, where it can be scaled, rotated, and repositioned. Users repeat this process across multiple source images to assemble a composition.} 
    \label{fig:user_interface-remixing-2d}
\end{figure}

\begin{figure}
    \centering
    \includegraphics[width=\linewidth]{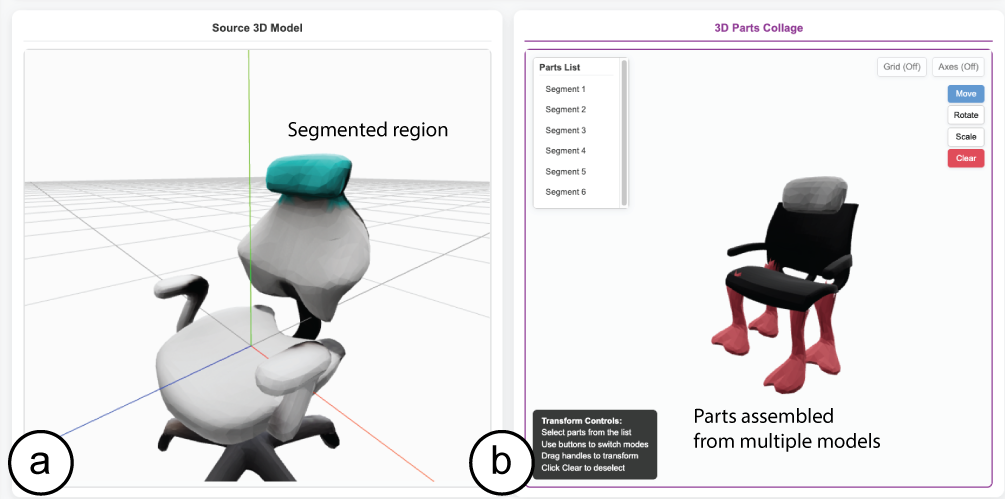}
    \caption{3D Remixing Workspace. (a)~The user selects a region on the source 3D model using click-based segmentation; the highlighted region (teal) shows the selected part. (b)~Extracted segments from multiple source models are assembled in the 3D parts collage, where each part can be individually moved, rotated, and scaled. A parts list tracks all segments.}
    \label{fig:user_interface-remixing-3d}
\end{figure}

\subsection{Part Selection}
Once the user is satisfied with the models added to the `Remix Context', they move to the Remixing page. Here, users segment and extract 3D parts for composition. Compos3D supports interactive segmentation via 2D images or 3D meshes, enabling fine-grained selection of regions to be carried forward into composition. 
\begin{figure}
    \centering
    \includegraphics[width=\linewidth]{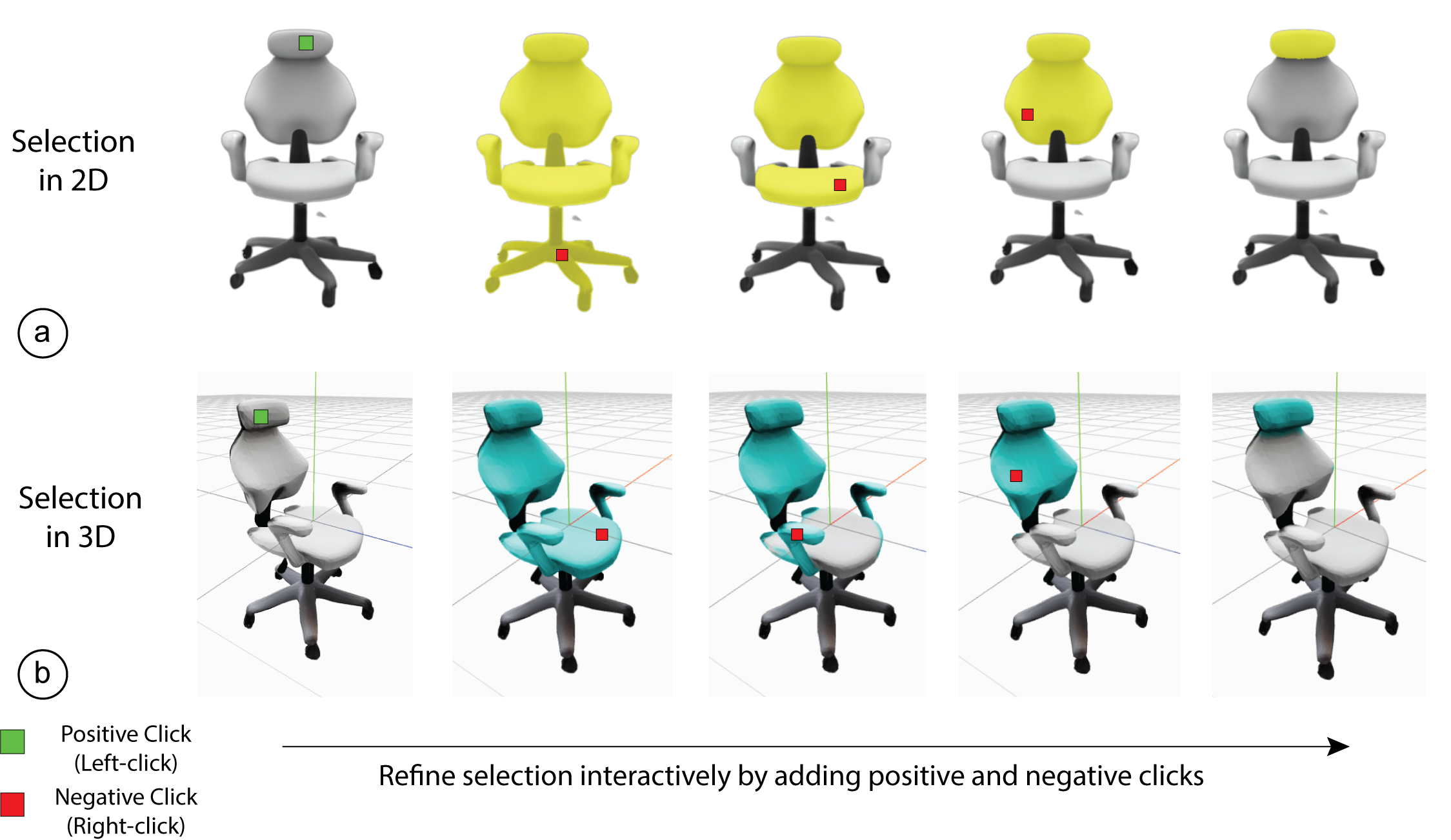}
    \caption{Interactive Part Selection in Compos3D. Users can segment regions either from (a) 2D images or from (b) 3D meshes. Both modalities support positive (green) and negative (red) clicks to iteratively refine selections until only the desired part is highlighted. Once finalized, selected regions are added to the remixing context for composition.}
    \label{fig:system_selection}
\end{figure}

\changes{
For part selection, we adopt a click-based interactive segmentation approach using the Segment Anything Model (SAM)~\cite{kirillov2023segment} for 2D images and Point-SAM~\cite{zhou2024point} for 3D meshes (Fig.~\ref{fig:system_selection}). We chose click-based prompting over alternatives such as brush selection, lasso tools, or sketch-based input because it offers the lowest interaction threshold for novice users: a single positive click (\texttt{left-click}) can identify a semantically meaningful region, and the selection can be iteratively refined by adding further positive clicks to expand the region or negative clicks (\texttt{right-click}) to exclude unwanted areas. This positive/negative prompting paradigm, introduced in SAM~\cite{kirillov2023segment}, supports both coarse semantic selections and finer boundary adjustments without requiring users to trace precise contours which can be challenging on 3D surfaces viewed through a 2D viewport.
}

In the 2D interface, users operate on individual images (Fig.~\ref{fig:user_interface-remixing-2d}). Users load an image, apply positive and negative clicks to adjust the highlighted region, view feedback in real time (Fig.~\ref{fig:user_interface-remixing-2d}b, ~\ref{fig:system_selection}a), and finalize the segment before adding it to the collage (Fig.~\ref{fig:user_interface-remixing-2d}c). This process can be repeated across multiple images to collect parts for remixing.

In the 3D interface, users interact directly with meshes in a 3D viewer. Point-SAM~\cite{zhou2024point} extends the positive/negative click-based prompting paradigm to 3D models. Highlighted regions are rendered directly on the mesh surface in real time (Fig.~\ref{fig:user_interface-remixing-3d}a, \ref{fig:system_selection}b). Once refined, the selected region is extracted as a mesh and added to the collage for later composition (Fig.~\ref{fig:user_interface-remixing-3d}b).

\subsection{Composition}

\begin{figure}
    \centering
    \includegraphics[width=\linewidth]{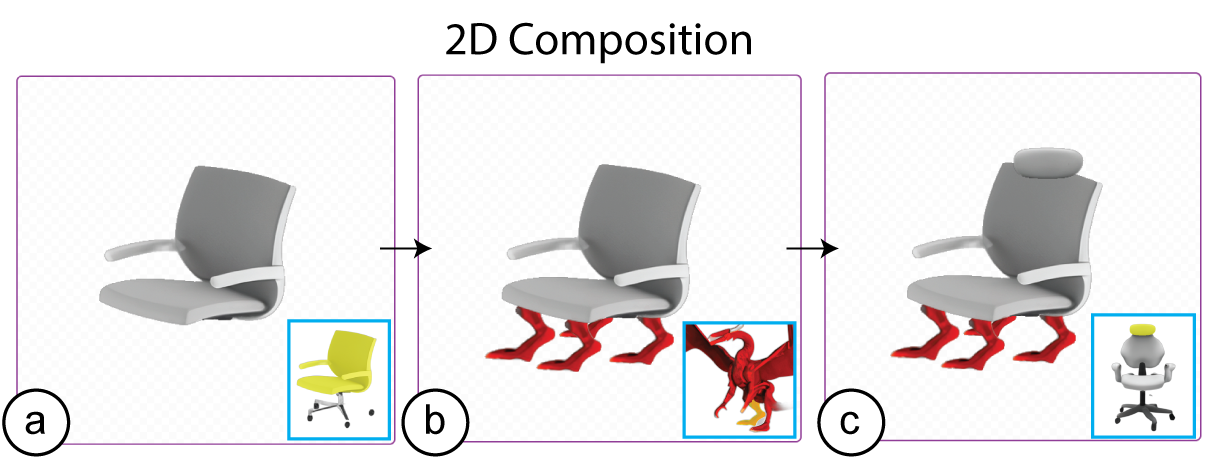}
    \caption{2D Composition. Users iteratively build a collage by selecting regions from source models (insets) and placing them on a 2D canvas. (a)~A chair seat extracted from one model. (b)~Dragon legs added from a second model. (c)~A headrest added from a third model. Elements can be scaled, rotated, and overlapped at each step.}
    \label{fig:system_composition-2d}
\end{figure}

\begin{figure}
    \centering
    \includegraphics[width=\linewidth]{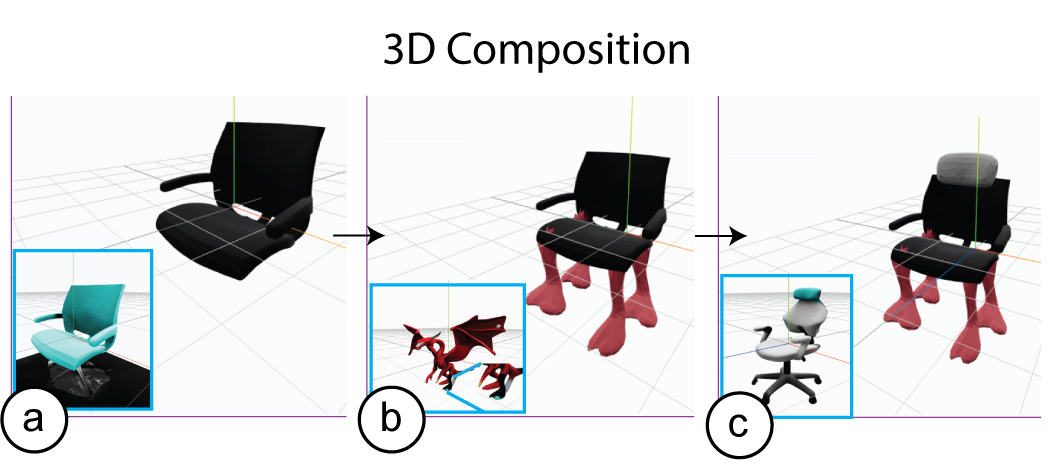}
    \caption{3D Composition. Users extract parts from source models (insets) and arrange them as meshes in 3D space. (a)~A chair seat placed in the scene. (b)~Dragon legs added and positioned. (c)~A headrest added and aligned. Parts can be translated, rotated, and scaled for precise spatial arrangement.}
    \label{fig:system_composition-3d}
\end{figure}

Once parts are selected and added, the system provides a canvas for composition, where users can explore different design possibilities by combining elements into new forms. This stage functions similarly to a canvas in creative practice, where designers assemble segments to explore ideas, compare alternatives, and evaluate new directions. Using the freeform canvas, users can visually experiment with the structure and style of their composition before finalizing the model in the synthesis stage.

In the 2D canvas, extracted image segments are placed as independent movable elements. Users can scale, rotate, and overlap them to explore alternative collages that represent their design intent. If they find a certain element that is missing, they can add additional parts from other images (Fig.~\ref{fig:system_composition-2d}). 

In the 3D canvas, the previously added segments appear as independent 3D meshes with the original material properties. Users can rotate, translate, and scale these parts individually in the 3D space to semantically align them according to their design intent (Fig.~\ref{fig:system_composition-3d}). Unlike the flat compositional space in 2D, the 3D canvas affords precise control over spatial arrangement, allowing users to resolve perspective and proportion as desired. 

\subsection{Synthesis}

\begin{figure}
    \centering
    \includegraphics[width=\linewidth]{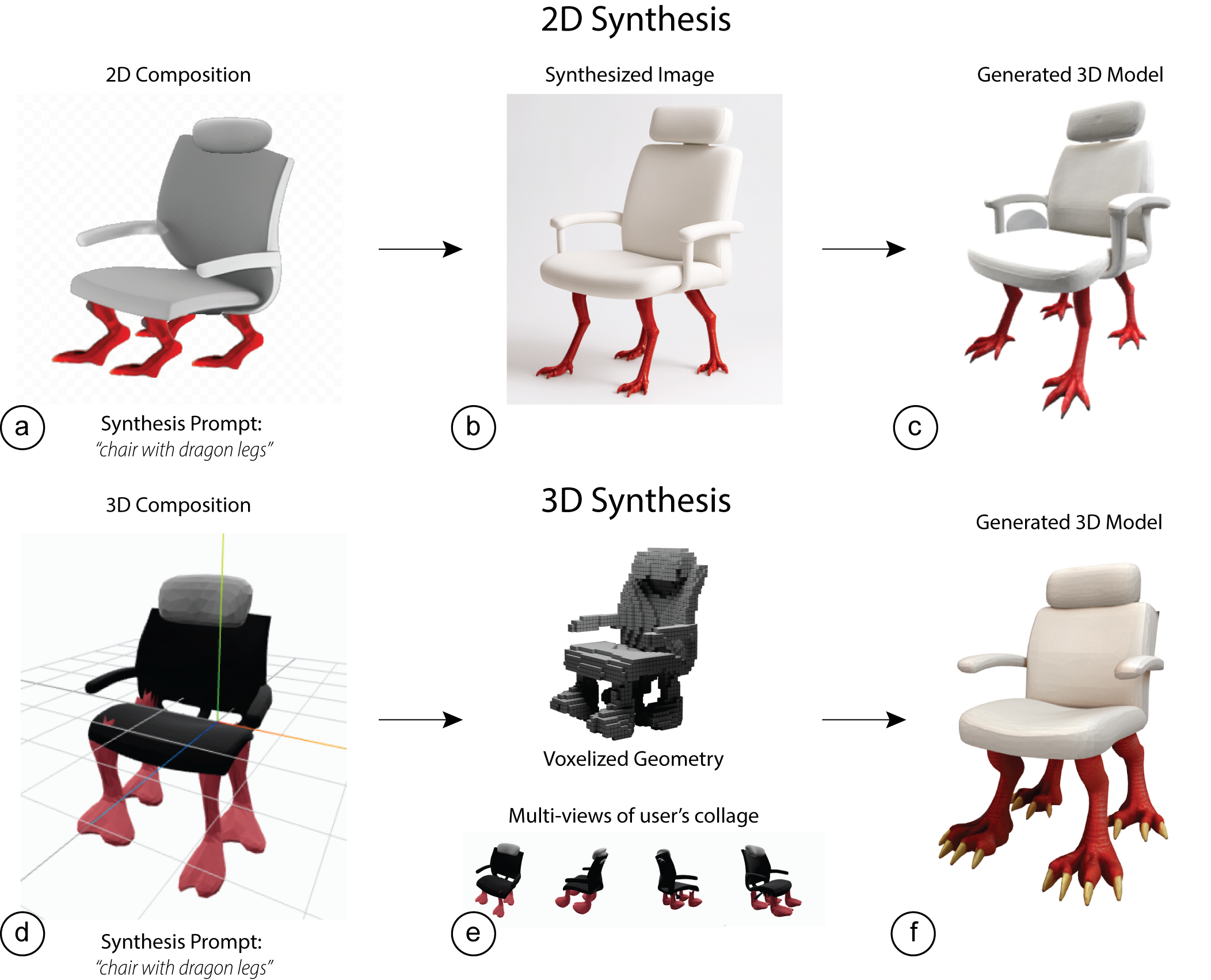}
    \caption{Synthesis in Compos3D. (Top)~2D synthesis: the user's collage is refined into a coherent image, then converted to a 3D model. (Bottom)~3D synthesis: the user's mesh composition is voxelized and decoded with multi-view features into a coherent 3D model. Both workflows accept an optional text prompt to guide synthesis.}
    
    \label{fig:system_synthesis}
\end{figure}

In this final stage, Compos3D transforms the rough compositions into coherent designs. This freeform compositing in the `Composition' phase allows users to experiment and clarify their design intent, producing a rough approximation of the intended structure and style. But since these segments are eclectically sourced from different pre-generated models and positioned manually by the user, these rough compositions may lack the coherent topological details required in a final mesh. The synthesis stage resolves this gap, connecting the 3D parts together to form watertight geometry and preserve surface detail.

For 2D synthesis, Compos3D processes the rough collage created by the user into a refined composite image, using an image-generation model. Optionally, the user can add a text prompt to provide additional detail (Fig.~\ref{fig:system_synthesis}a). The system treats the collage as a structural guide, producing coherent images that retain the composition but add detail and consistency. If the user wishes to further update the collage, they can optionally regenerate alternatives until the result reflects their design intent. The finalized image is then converted into a 3D model through an image-to-3D pipeline. 

For 3D synthesis, Compos3D first voxelizes the rough collage assembled by the user, producing a structured volumetric scaffold that encodes the overall form and proportions of the design. In parallel, the system renders multiple views of the collage and refines them with an image-to-image model (optionally guided by user text prompts), ensuring that the synthesized appearance reflects the intended structure and semantic details. These multi-view images, together with the voxelized geometry, are then passed into the TRELLIS pipeline~\cite{trellis}, which operates on a Structured Latent (SLAT) representation to jointly generate geometry and appearance. By fixing the voxel structure, and combining with dense multi-view features, Compos3D uses TRELLIS to generate a coherent 3D mesh that preserves the high-level composition specified by the user while resolving seams, topology, and surface detail. The refined 3D model is then presented back to the user.

\subsection{Technical Implementation}

Compos3D integrates multiple generative AI models into a modular pipeline for segmentation, composition, and synthesis. Given the rapid progress in generative AI, we designed the system so that any component can be replaced with an improved model without changing the workflow. We use TRELLIS~\cite{trellis} as our core 3D generation backbone, as it is the current state-of-the-art model trained on the large-scale Objaverse dataset~\cite{deitke2023objaverse}, enabling generalized performance across diverse object categories.

For 2D synthesis, the user's collage is composited into a single image and refined using an image generation model (OpenAI's gpt-image-1), optionally guided by a user text prompt. The refined image is then passed to the TRELLIS image-to-3D pipeline, which predicts consistent multi-view renderings and reconstructs a 3D mesh.

For 3D synthesis, we leverage the internal structure of TRELLIS to preserve the user's spatial composition. When users segment each mesh in the remixing workspace, we track the mesh ID and the vertex mask from the original model. This allows us to identify the corresponding voxels in TRELLIS's intermediate Structured Latent (SLAT) representation, where each voxel encodes both geometric and appearance features. As users manipulate segments in the 3D canvas (translation, rotation, and scale), we record these transformations as homogeneous coordinate matrices for each part. During synthesis, we apply these same transformations to the corresponding voxels in the SLAT representation, producing a composite latent volume that reflects the user's spatial arrangement. In parallel, the system renders multiple 2D views of the user's 3D collage and refines them with image-to-image generation, optionally guided by user text, producing multi-view features that capture the intended appearance. This composite latent volume is then decoded together with these multi-view features into a coherent 3D mesh using TRELLIS's decoder network, which resolves seams, topology, and surface detail while preserving the high-level structure specified by the user. On an NVIDIA L4 GPU, end-to-end synthesis runs in an average of 42.3 seconds (SD = 7.2s) per model, enabling interactive generation.

\section{User Study}
\label{sec:user_study}


To evaluate whether our compositional remixing offers distinct benefits over the regeneration workflow, and to compare 2D and 3D interaction modalities, we conducted a controlled user study centered around three research questions:


\begin{enumerate}
    \item \textbf{RQ1 (Workflow Preference):} How does compositional remixing compare with the regeneration workflow in terms of creative control, expressiveness, and alignment with user intent?
    \item \textbf{RQ2 (Modality Preference):} How do 2D and 3D interaction modalities compare in terms of usability, effort, and perceived effectiveness within each workflow?
    \item \textbf{RQ3 (Trade-offs):} What trade-offs do users perceive across workflows and modalities in terms of effort, control, and satisfaction?
    
\end{enumerate}

In the following subsections, we describe the study design and baseline workflow, participant demographics, and the tasks and procedure used in the evaluation.

\subsection{Study Design \& Baseline Workflow}

The study followed a within-subjects $2 \times 2$ factorial design with two independent variables: \textbf{Workflow} (Remixing vs. Regeneration) and \textbf{Modality} (2D image-based vs. 3D segment-based interaction). Each participant completed four conditions: Remixing–2D, Remixing–3D, Regeneration–2D, and Regeneration–3D. Task order was counterbalanced using a Latin Square. With eight participants, we completed two full rotations of the Latin Square, ensuring that each condition appeared equally often in each order position. The study duration was 2 hours.



\begin{figure*}
    \centering
    \includegraphics[width=\linewidth]{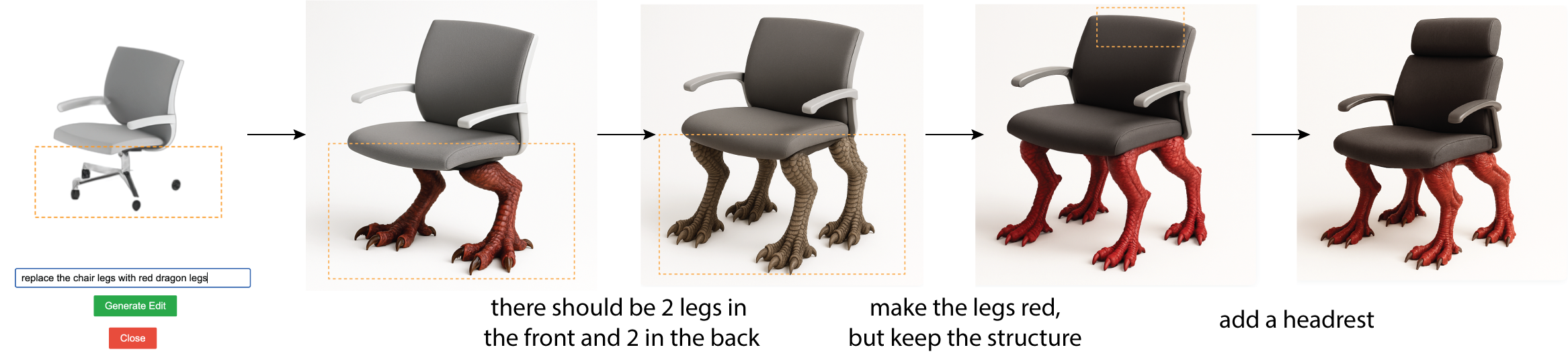}
    \caption{Regeneration (baseline) workflow in Compos3D. Users highlight a local region (via bounding box in 2D or surface selection in 3D) and provide a textual instruction to specify the desired change. The system regenerates the highlighted region while preserving the rest of the design, allowing users to progressively refine their model across multiple iterations.}
    \label{fig:iterative_workflow}
\end{figure*}

\paragraph{Baseline: Regeneration Workflow.} As a baseline, we implemented a \textit{regeneration workflow} in both 2D and 3D (Fig~\ref{fig:iterative_workflow}), mirroring the dominant interaction paradigm in current 3D generative tools such as Meshy AI~\cite{meshy2026} and Hunyuan3d~\cite{zhao2025hunyuan3d}. To ensure a fair comparison, this workflow was implemented within the Compos3D framework, sharing the same interface, generation pipeline, and model quality.

\changes{
In this workflow, users generate an initial 3D model from a text or image prompt as described in Section~\ref{sec:initial_model_gen} and regenerate undesired regions, akin to a dice roll. Specifically, users select a local region of interest (bounding box in 2D or surface click in 3D) and provide a textual instruction describing the desired edit (e.g., ``replace legs with dragon legs''), as depicted in Figure~\ref{fig:iterative_workflow}. The system regenerates the model with the context of the user's input, and users repeat this cycle until satisfied. The key distinction from remixing is that regeneration requires users to refine their design through local text-guided edits, relying on the model to interpret each instruction. In contrast, remixing provides more fine-grained control over the final outcome, allowing users to compose the design directly from multiple source models.
}


\subsection{Participants}

We recruited eight participants (4 male, 4 female), aged 21–39 years (M = 27.0, SD = 5.76), through snowball sampling and institute mailing lists. All participants were beginners with 3D modeling tools. Participants’ familiarity with AI generative tools (e.g., ChatGPT, DALL-E) was moderate (M = 4.38, SD = 1.60, on a 7-point scale), ranging from beginner (1) to expert (7). The entire user study lasted for 2 hours, and the participants were provided with a 120 USD gift card for their time.


\subsection{Tasks and Procedure}

Each participant completed four design tasks, one in each condition of the $2 \times 2$ design (Remixing–2D, Remixing–3D, Regeneration–2D, Regeneration–3D). To allow direct within-subject comparisons while holding task requirements constant, all tasks were based on the \emph{same} design brief:

\begin{quote}
\textbf{Design a hybrid work chair.}\\
\textit{Backrest:} Tall, arched frame.\\
\textit{Armrests:} Jointed supports as branches, with one armrest featuring a swing-out tray console and control screen.\\
\textit{Base:} Rotating foundation with a footrest platform in the front.\\
\textit{Headrest:} Adjustable head support with lighting.\\
\textit{Seat Cushion:} Lattice structure with a soft material.
\end{quote}

A design brief provided a controlled task, where functional and aesthetic requirements are held constant across conditions, while allowing for creative interpretation and freedom. Design briefs also mirror real-world design practice, where requirements containing both function and style are specified as written descriptions~\cite{hegemann2024palette, cross2006designerly}. 

At the start of each session, participants completed a background questionnaire on demographics, prior 3D modeling experience, familiarity with generative AI tools, and exposure to remixing workflows. They then received a short tutorial covering all four conditions and completed a warm-up design task using a simplified 3-part brief (designing a mug) to familiarize themselves with the interface.

Each participant then completed the four experimental conditions in counterbalanced order using a Latin Square. In each condition, they (1) reviewed the design brief, (2) completed the modeling task with the assigned tool and modality, and (3) answered a post-task questionnaire rating task outcome, perceived control, cognitive demand/effort, enjoyment, and tool responsiveness, with optional open-ended feedback.  

After completing all tasks, participants filled out a post-study questionnaire on overall preferences (tool and modality), perceived control, and confidence in meeting the design brief. Finally, we conducted a semi-structured interview to capture deeper reflections on creative strategies, sense of control, and envisioned roles for generative AI in future design practice.



\section{Results}
\changes{To compare the workflows, we measured both quantitative and qualitative outcomes. Likert-scale ratings from post-task questionnaires were analyzed per item. For each comparison, we averaged each participant's responses across the other factor: across modalities for the workflow comparison (RQ1), and across tools for the modality comparison (RQ2), yielding paired scores ($n=8$) for each survey question. We report sample means with 95\% confidence intervals, computed using Student's $t$ distribution. We support these quantitative patterns with qualitative evidence from open-ended responses and interviews, which were analyzed using Reflexive Thematic Analysis~\cite{braun2006using, braun2019reflecting} to capture recurring strategies, perceived strengths and limitations, and reflections on creative control.}

\subsection{RQ1: Remixing vs. Regeneration}

\begin{figure}[t!]
    \centering
    \includegraphics[width=\linewidth]{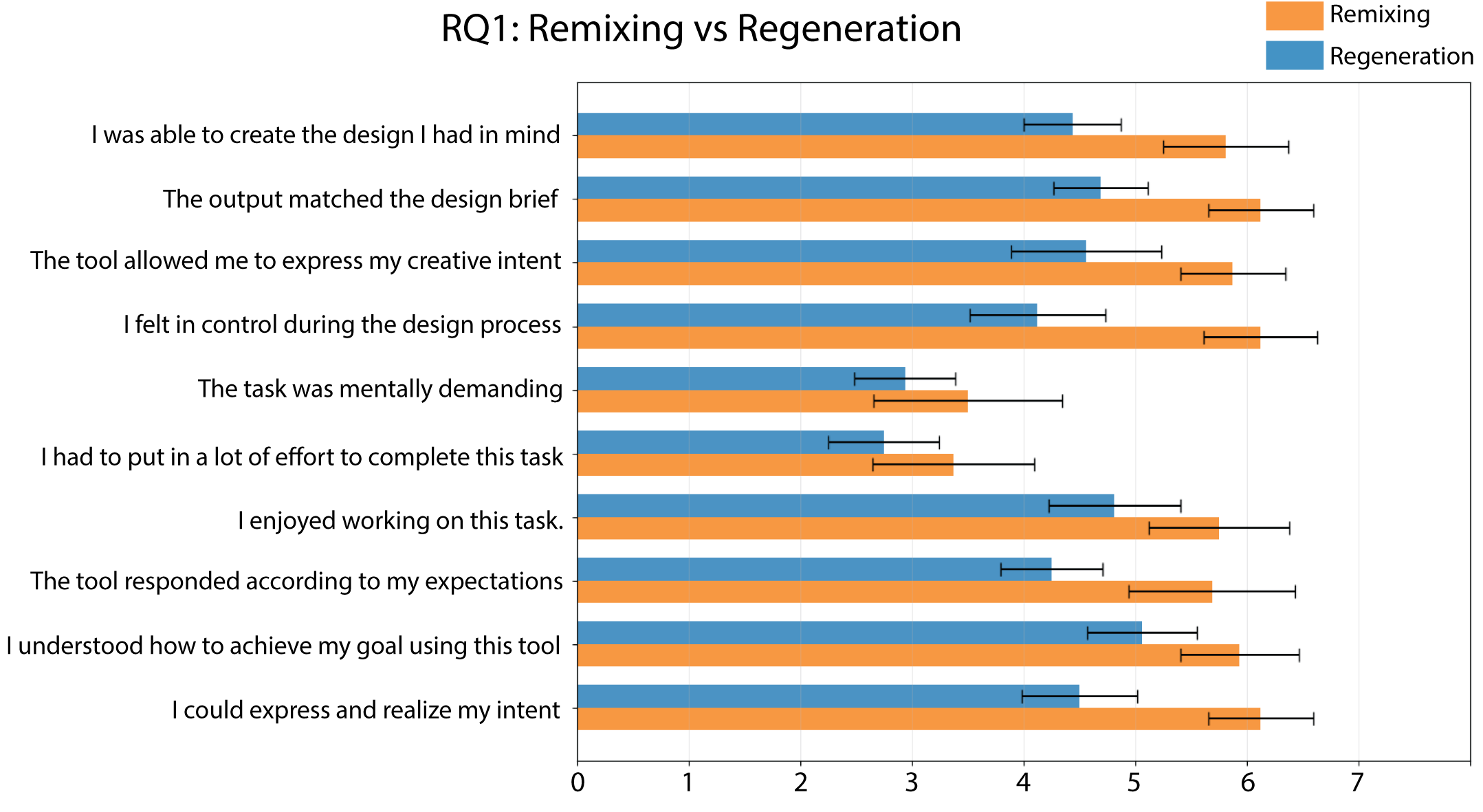}
    \caption{Mean participant ratings (Likert 1–7; 95\% CIs) for \textbf{Remixing} (orange) vs.\ \textbf{Regeneration} (blue) across \emph{Task Outcome} (Q1–Q4), \emph{Cognitive Demand \& Effort} (Q5–Q7), and \emph{Tool Responsiveness} (Q8–Q10). Remixing trends higher on creative expression, perceived control, and responsiveness, with a modest increase in reported effort.
    }
    \label{fig:user_study_remix_vs_regen}
\end{figure}

To compare the two workflows, we analyzed participants’ responses to the post-task questionnaire across three dimensions: \emph{Task Outcome} (Q1–Q4), \emph{Cognitive Demand \& Effort} (Q5–Q7), and \emph{Tool Responsiveness} (Q8–Q10). Figure~\ref{fig:user_study_remix_vs_regen} summarizes these quantitative results.


\textbf{Task Outcome.} Participants rated Remixing consistently higher than Regeneration across all outcome measures. They more strongly agreed that they were able to create the design they had in mind (Q1: Remix $M=5.81$, $SD=1.05$ vs.\ Regeneration $M=4.44$, $SD=0.81$) and that the output matched the design brief (Q2: Remix $M=6.12$, $SD=0.89$ vs.\ Regeneration $M=4.69$, $SD=0.79$). They also reported greater support for creative expression (Q3: Remix $M=5.88$, $SD=0.89$ vs.\ Regeneration $M=4.56$, $SD=1.26$) and felt more in control of the design process (Q4: Remix $M=6.12$, $SD=0.96$ vs.\ Regeneration $M=4.12$, $SD=1.15$).  

\textbf{Cognitive Demand \& Effort.} Ratings here were more mixed. While participants reported slightly higher mental demand and effort in the Remixing condition (Q5: $M=3.50$, $SD=1.59$ vs.\ Regeneration $M=2.94$, $SD=0.85$; Q6: $M=3.38$, $SD=1.36$ vs.\ $M=2.75$, $SD=0.93$), they nonetheless found Remixing more enjoyable (Q7: $M=5.75$, $SD=1.18$ vs.\ $M=4.81$, $SD=1.11$). This suggests that the increased effort of Remixing was offset by greater satisfaction and engagement.  

\textbf{Tool Responsiveness.} Participants rated Remixing as significantly more responsive and aligned with their goals. They reported that the tool responded according to expectations (Q8: Remix $M=5.69$, $SD=1.40$ vs.\ Regeneration $M=4.25$, $SD=0.86$), that they understood how to achieve their goals more clearly (Q9: Remix $M=5.94$, $SD=1.00$ vs.\ Regeneration $M=5.06$, $SD=0.93$), and that they could express their intent more effectively (Q10: Remix $M=6.12$, $SD=0.89$ vs.\ Regeneration $M=4.50$, $SD=0.97$).  

Taken together, these results indicate a consistent advantage for Remixing: despite requiring slightly more effort, it provided participants with greater creative control, stronger alignment with their design intent, and a more enjoyable overall experience.

\paragraph{Qualitative Findings:} 
Participants consistently emphasized that \emph{Remixing} gave them a stronger sense of creative control than Regeneration. Several noted that Regeneration often removed or altered features they wanted to keep: ``\textit{Regeneration sometimes removes features I want, but remixing gives me an option to bring those lost features back into the final model}'' (P6). Others framed their preference in terms of reduced unintended changes and explicit control over the result: ``\textit{It's hard to generate an image or a 3D model that is perfect, and remixing allows me to combine different parts which I think are perfect in their own setting. It makes me feel I am controlling the process}'' (P3). This sense of control extended beyond feature preservation to spatial manipulation, with P4 noting ``\textit{I feel in more control on objects, including the position and scale}.'' Overall, participants described Remixing as a more intentional and reliable process.

\subsection{RQ2: 2D vs. 3D Modality}
\label{sec:rq2}

\begin{figure}[t!]
    \includegraphics[width=\linewidth]{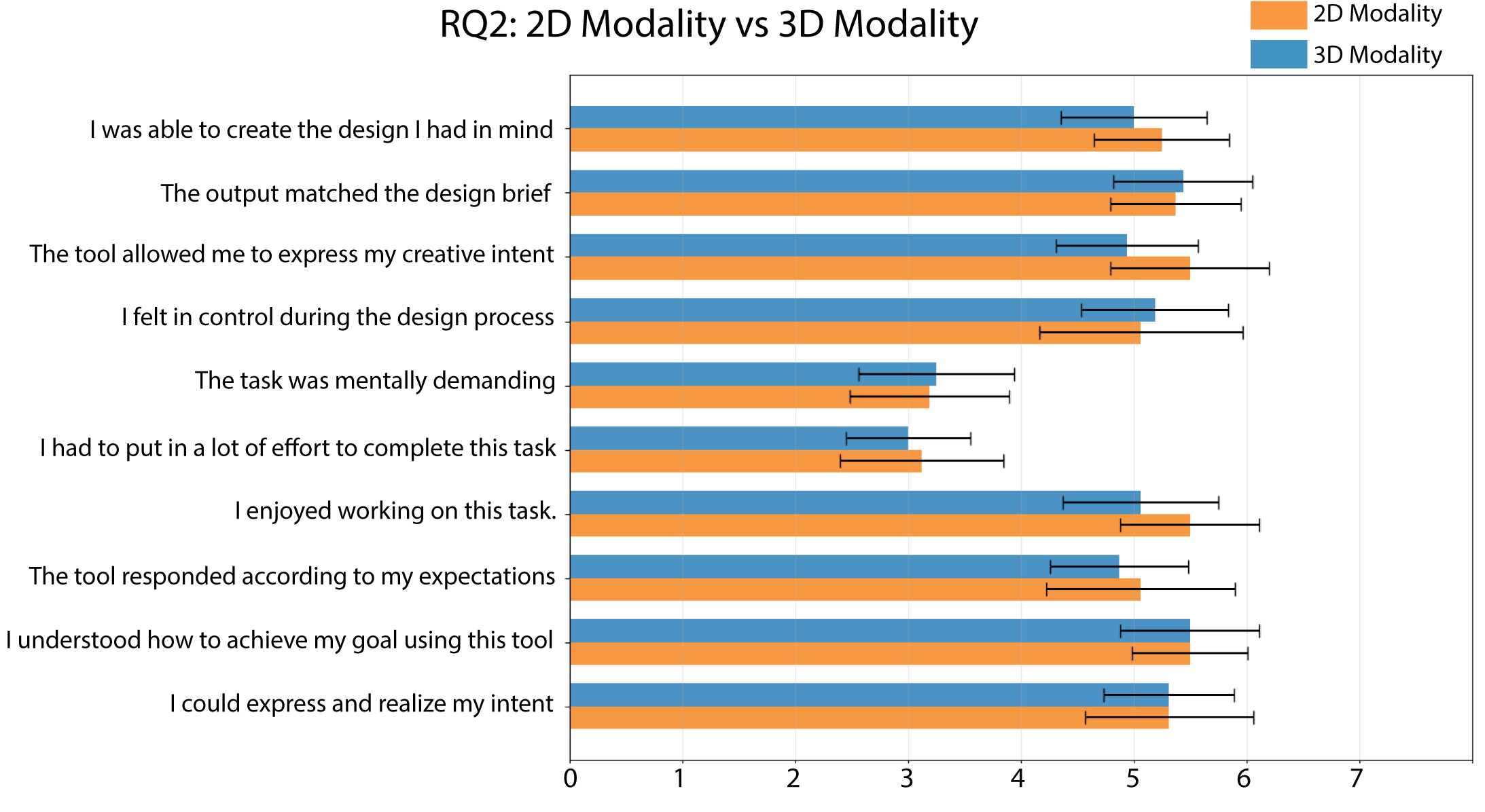}
    \caption{Mean participant ratings (Likert 1–7; 95\% CIs) for \textbf{2D} (orange) vs.\ \textbf{3D} (blue) across \emph{Task Outcome} (Q1–Q4), \emph{Cognitive Demand \& Effort} (Q5–Q7), and \emph{Tool Responsiveness} (Q8–Q10). Ratings were broadly similar across modalities.
}
    \label{fig:user_study_2d_vs_3d}
\end{figure}

To compare the two modalities, we analyzed post-task questionnaire responses across the same three dimensions (Figure~\ref{fig:user_study_2d_vs_3d}). Ratings for 2D and 3D were largely similar across all measures. The 3D modality was rated somewhat higher for supporting creative expression (Q3: 3D $M=5.50$, $SD=1.32$ vs.\ 2D $M=4.94$, $SD=1.18$) and enjoyment (Q7: 3D $M=5.50$, $SD=1.15$ vs.\ 2D $M=5.06$, $SD=1.29$), while task outcome, cognitive demand, and tool responsiveness ratings were comparable across modalities.

\paragraph{Qualitative Findings.}
Most participants reported that the 3D modality felt more intuitive because it allowed direct spatial manipulation without mental translation: ``\textit{Working with 3D was more intuitive than working with 2D because when working with 2D I always had to think about how the 2D–3D translation will affect the shape and design}'' (P5). At the same time, a minority found 2D interactions simpler and less ambiguous, particularly for segmentation: ``\textit{I think the selection aspect of 2D was more intuitive than the view/angle part of 3D, and it seemed to respond better}'' (P7). Overall, participants valued 3D for precision and spatial reasoning, but recognized that 2D offered speed and simplicity in selection.

\subsection{RQ3: Trade-offs Across Conditions}

To complement the per-task ratings, the post-study questionnaire asked participants to directly compare the four conditions. Participants unanimously preferred Remixing over Regeneration (8/8), and most found 3D more intuitive (6/8). Remixing–3D was most often selected as the most accurate (5/8), allowing the most creative freedom (6/8), and the condition in which participants felt most confident meeting the brief (5/8). Effort was most strongly associated with Regeneration–3D (5/8), followed by Remixing–3D (3/8). This highlights a clear trade-off: Remixing–3D provided the greatest control and confidence but also demanded 
more effort, while 2D workflows were seen as faster and more straightforward but less controllable.

\subsection{Exploratory Reflections}   
In addition to addressing our research questions, we asked participants open-ended questions about their design strategies, attachment to outcomes, and perspectives on how AI-based 3D modeling tools should evolve.

\textbf{Interactive engagement is valued even if AI were `perfect'.}  
All participants (8/8) reported that they would still want to interactively design even if AI could generate a flawless result from a single prompt. They stressed that design intent often emerges during the process: \textit{``I don't have a very clear idea of what I want at the start, and I need the process to find it out while designing''} (P1). Others emphasized that prompting alone felt limiting: ``\textit{A prompt always leaves a lot open in regards to execution}'' (P3), and ``\textit{Creativity cannot be expressed in one step, but only in multiple iterations}'' (P7).

\textbf{Design thinking alternates between whole objects and parts.}  
Participants described shifting fluidly between reasoning about the whole design and its individual components. For some, part-based thinking was primary: ``\textit{When designing this model, I was definitely thinking more in terms of parts or components}'' (P5). Others alternated: ``\textit{I switched between parts and the whole a lot of times}'' (P1), while some found that whole-object thinking helped brainstorm before extracting parts: ``\textit{Thinking in terms of whole objects allowed me to brainstorm ideas and also extract interesting parts}'' (P7).

\textbf{2D and 3D modalities shape different modes of thinking.}  
Beyond the usability trade-offs reported in RQ2, participants reflected on how each modality affected their sense of authorship. P1 observed that ``\textit{In 2D, AI kind of takes over the design [\ldots] the results are nice but I don't feel as much that they are mine},'' whereas in 3D, ``\textit{I have very explicit control [\ldots] more effort but more precise results.}'' This suggests that the choice of modality influences not just efficiency but the perceived creative agency.

\textbf{Ownership emerges from composing parts.}  
Nearly all participants stated they felt more attached to designs composed from parts than to those generated all at once: \textit{``Generating didn't feel like I was creating, more like typing something and seeing what happens''} (P2). P5 shared that composition led to ownership: \textit{``I did feel more attached to designs composed from parts... [as I] need to think about which parts to include and how to arrange them.''} 

Together, these reflections suggest that AI-assisted 3D modeling tools should prioritize \textit{co-creation over automation}, supporting workflows where users iteratively shape, refine, and compose designs.

\section{Applications}


To illustrate use cases for Compos3D beyond the study task, we present three applications that highlight how remixing workflows extend prompt-based generation: assembling designs from multiple sources, reframing generative flaws as opportunities for creative repair, and composing personal artifacts into new fabricable designs.

\subsection{Creating Out-of-Distribution 3D Designs with Generative AI}

\begin{figure}
    \centering
    \includegraphics[width=\linewidth]{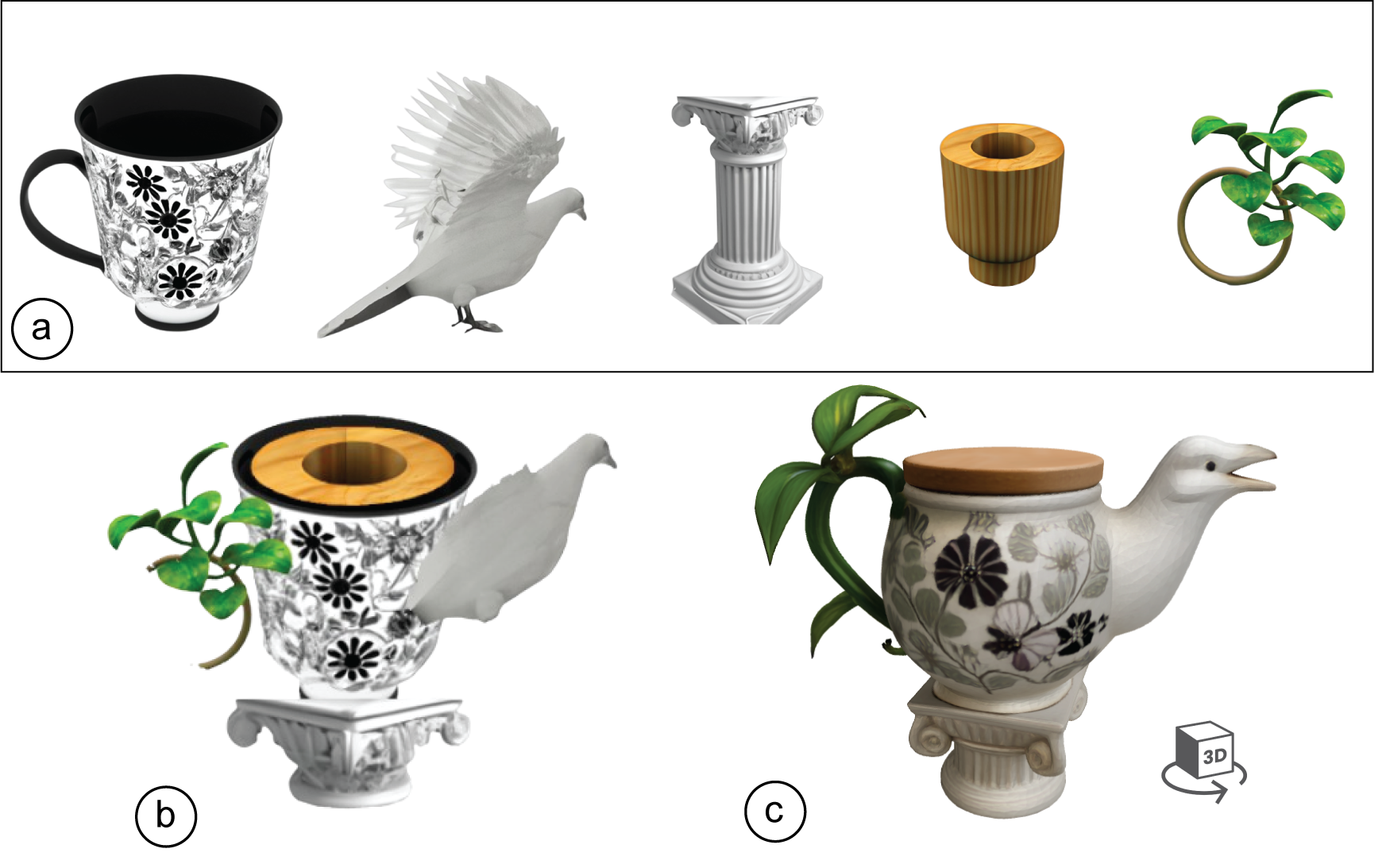}
    \caption{Creating Out-of-Distribution 3D Designs. (a)~Source components from multiple generated models. (b)~User-assembled collage. (c)~Synthesized teapot integrating the parts into a coherent design.}
    \label{fig:app_composition_figure}
\end{figure}



\changes{Remixing enables users to create 3D designs that go beyond what any single generation can produce. These designs can include unconventional forms that are \textit{out-of-distribution}, i.e., unlikely to appear in the model’s training data. By combining elements across multiple results, remixing expands users’ creative capabilities and allows them to construct novel objects that would be difficult to generate with a single prompt. In Figure~\ref{fig:app_composition_figure}, a user assembled five separately generated components: a floral motif mug, dove spout, vine handle, Roman pedestal base, and bamboo lid, into a collage (Fig.~\ref{fig:app_composition_figure}b). Compos3D synthesized them into a coherent teapot (Fig.~\ref{fig:app_composition_figure}c) that preserves the recognizable form of each part while maintaining overall aesthetic. The resulting design lies outside the distribution of typical designs, illustrating how remixing enables highly novel designs.}


\subsection{Remixing as Repair}

\begin{figure}
    \centering
    \includegraphics[width=\linewidth]{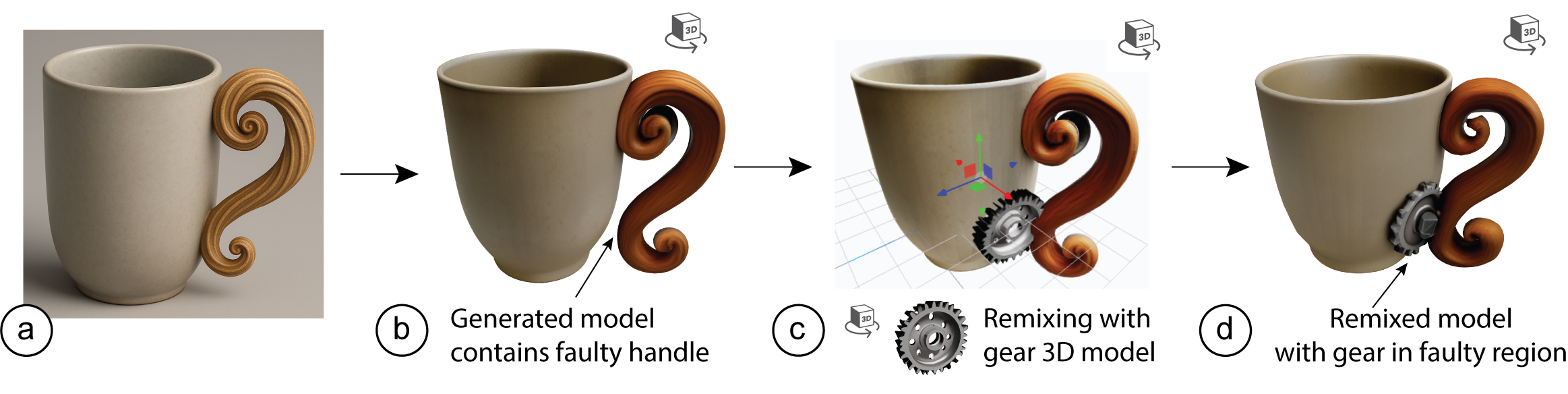}

    \caption{Remixing as repair. (a)~A generated mug image. (b)~The resulting 3D model contains a faulty handle. (c)~A gear model is inserted into the faulty region. (d)~The synthesized result integrates the gear as a distinctive design feature.}
    \label{fig:remixing_as_repair}
\end{figure}

Generative 3D modeling may produce incomplete or broken geometry that requires post-processing or refinement~\cite{trellis}. 
While users can attempt to fix these flaws through regeneration or manual correction, compositional remixing offers an additional option for creative intervention.
Inspired by Japanese \textit{Kintsugi}~\cite{santini2019kintsugi}, where broken pottery is repaired with precious materials to highlight rather than conceal damage, Compos3D enables users to treat errors as sites of creative expression. As shown in Figure~\ref{fig:remixing_as_repair}, a generated mug with a faulty handle is repaired by inserting a gear model, which the system synthesizes into a distinctive design feature, transforming a defect into an intentional element without requiring low-level 3D modeling operations.

\begin{figure}
    \centering
    \includegraphics[width=\linewidth]{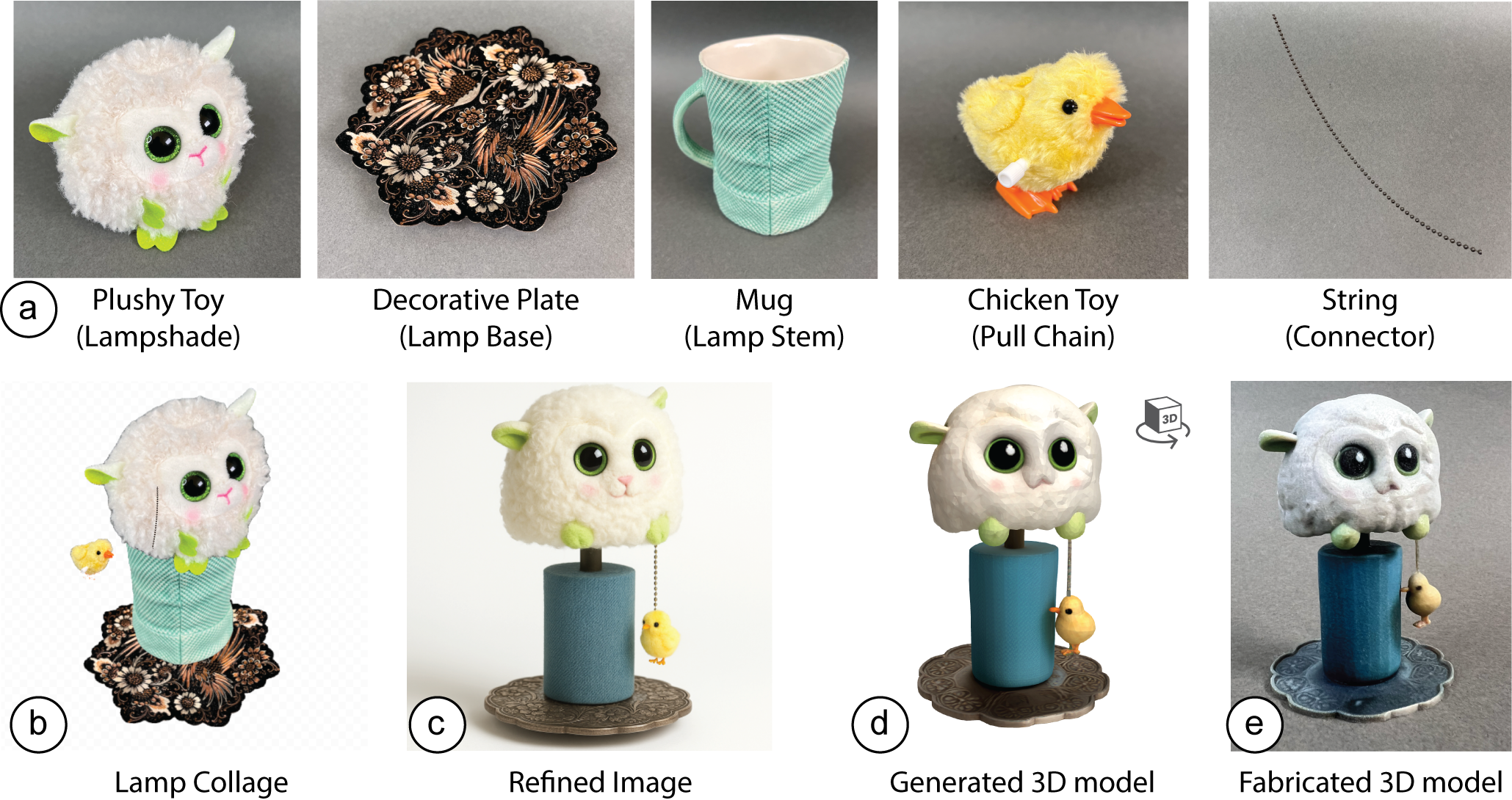}
    \caption{Composing personal artifacts. (a)~Everyday items used as source parts. (b)~Assembled collage. (c)~Refined composite image. (d)~Generated 3D model. (e)~Fabricated lamp.}
    \label{fig:existing_artifacts}
\end{figure}

\changes{
\subsection{Composing Personal Artifacts for Fabrication}

Users often associate strong memories and personal meaning with everyday artifacts such as gifts, souvenirs, or toys, that they want to preserve and reimagine in new contexts. Prior work on memory reconstruction, such as InteRecon~\cite{li2025interecon}, highlights how digitizing personal items can serve as lasting memory triggers, but existing methods typically replicate objects as static forms. With Compos3D, users can upload photos of personal artifacts, recreate them as generative 3D models, and remix them into new compositions, enabling objects to live on as active design materials rather than preserved scans. We demonstrate this in Figure~\ref{fig:existing_artifacts} with a lamp assembled from everyday items: a plush toy as the lampshade, a decorative plate as the base, a mug as the stem, and a toy chicken as a pull chain. As shown in Figure~\ref{fig:existing_artifacts}e, the resulting design can also be fabricated for real-world use, closing the loop from physical inspiration to digital remixing to personal fabrication~\cite{baudisch2017personal, faruqi2023style2fab, faruqi2025tactstyle}.
}





\section{Discussion}


In this section, we translate our findings into design recommendations for future systems, discuss hybrid workflows, and outline limitations and directions for future work.

\subsection{Design Recommendations}

From our study findings and exploratory reflections, we distill two higher-level themes for future AI-assisted 3D modeling systems.


\textbf{1. Support Part-Based Composition in 3D Generation Workflows.}
Participants repeatedly valued the ability to shape outcomes through deliberate selection and arrangement rather than passive generation. As P3 reflected, ``\textit{even though AI is refining the final model, the process of composing myself and critiquing different elements made me feel like a designer}.'' P6 echoed that ``\textit{generating individual parts that can be composed felt more natural and intuitive}.'' We recommend treating generated 3D models as composable source material, supported by robust segmentation, flexible arrangement, and synthesis that preserves user-specified structure. Part-based composition can complement regeneration, allowing users to converge on desired design while generating new possibilities.


\textbf{2. Enable Fluid Switching Between 2D and 3D Interaction.}
Our modality comparison revealed that 2D and 3D interactions serve complementary roles, and that each shapes how users think about design. P4 observed, ``\textit{in 2D interactions I would more consider it as just a picture, not a 3D model. In 3D, I feel I have a completely different mindset}.'' P3 noted that ``\textit{during 2D interaction, you have to imagine how the design may look like from different perspectives, which is tiring. During 3D interaction, I can focus more on the different parts}.'' Rather than forcing users into one modality, we suggest future systems support fluid transitions; for example, selecting parts quickly in 2D and then refining their spatial arrangement in 3D, scaffolding laborious 3D tasks with aids such as alignment snapping and automated positioning.



Together, these themes highlight that the value of generative 3D design lies not only in producing high-quality outputs but in enabling workflows that balance efficiency, precision, and agency.

\paragraph{Hybrid Workflows: Combining Remixing and Regeneration}
While our study compared remixing and regeneration as distinct workflows for comparative analysis, our findings suggest that future systems could benefit from integrating them. Remixing affords deliberate preservation and recombination, while regeneration supports exploration of stylistic alternatives. An important opportunity for future work is to design hybrid workflows that allow users to fluidly move between the two within the same creative session; remixing supports preserving key parts, and regeneration allows exploration of broader diversity, supporting both precision and breadth in generative 3D modeling.

\subsection{Limitations and Future Work}
Our study has several limitations that suggest directions for future work. First, we evaluated a single multi-part design task with eight novice participants. Future work can explore a wider range of tasks and larger participant pools with users with varying levels of design expertise. Second, task outcomes were assessed through participant self-report. Future investigations can incorporate objective evaluations, such as independent expert ratings of design brief alignment to strengthen the evidence. Third, while our applications demonstrate iterative remixing (using outputs as inputs for further composition), our study evaluated only single-pass remixing. Future studies can investigate how iterative remixing shapes the design process over longer sessions. Finally, our current system recombines 3D geometry but does not account for fabrication constraints such as printability or material properties. Future work can integrate physics-aware generation and adaptive interfaces that recommend compositional edits based on user history.

\section{Conclusion}
In this work, we introduced Compos3D, a system that supports a compositional remixing workflow for 3D generation. 
Rather than treating generated models as final outputs to accept or regenerate, Compos3D treats them as intermediate source material from which users composite the final design.
Compos3D allows users to select desired parts from multiple generated models (via 2D or 3D modalities) and composite them into the final output. 
Our controlled user study demonstrates that this novel remixing-based approach requires slightly more effort, but provides greater creative control, alignment with user intent, and enjoyment compared to the traditional regeneration approach. We further highlight trade-offs between 2D and 3D modalities: while 2D offers speed and simplicity, 3D affords precision and spatial reasoning. From these findings, we distilled design guidelines that emphasize collaborative creation with AI, support for part-based thinking, and integration of multiple modalities.

\bibliographystyle{ACM-Reference-Format}
\bibliography{references}

\end{document}